\documentclass{article}
\usepackage{amssymb}
\usepackage{multirow}
\usepackage{hyperref} 
\usepackage{graphicx}

\title{A Neural Network Approach to ECG Denoising}

\author{Rui Rodrigues (rapr@fct.unl.pt) and Paula Couto (mpcc@fct.unl.pt)\\ {\small Departamento de Matem\'{a}tica da Faculdade de Ciencias e Tecnologia da UNL, 2829-516 Caparica, Portugal } }
\date{December 19,2012}

\begin{document}
\maketitle

\begin{abstract}
We propose an ECG denoising method based on a feed forward neural network with three hidden layers. Particulary useful for very noisy signals, this approach uses the available ECG channels to reconstruct a noisy channel. We tested the method, on all the records from Physionet  MIT-BIH Arrhythmia Database, adding electrode motion artifact noise. This denoising method improved the perfomance of publicly available ECG analysis programs on noisy ECG signals. This is an offline method that can be used to remove  noise from very corrupted Holter records.
\end{abstract}

\section{Introduction}

 The ECG is often corrupted by different types of noise, namely, power line interference, electrode contact and motion artifacts, respiration, electrical activity of muscles in the vicinity of the electrodes and interference from other electronic devices. Analysis of noisy ECGs is difficult for humans and for computer programs. 
In this work we place ourselves in context of automatic and semi automatic ECG analysis: denoising should facilitate  automatic ECG analysis.

General denoising signal processing methods  have been applied to ECG. Low pass linear filters are used for high frequency noise removal, namely power line interference and muscle activity artifacts. High pass linear filters can be applied to cancel  baseline wander.  The use of neural networks to ECG denoising has been, to our knowledge, limited to the removal of these two types of noise. Other denoising tools are median filter, wavelet transform methods, empirical mode decomposition, morphological filters, non linear bayesian filtering  and  template matching. We will focus on noise introduced by electrode motion which causes more difficulties in ECG analysis\cite{Moodynoisestress}.
 Our method adapts to each particular ECG channel and learns how to reproduce it from a noisy version of the different channels available.

In the Physionet/Cinc Challenge 2010 it was shown that we can use some physiological signals to reconstruct another physiological signal, in particular an ECG~\cite{Moody2010,fillinginthegap, missingphysiological}. 
 This approach to reconstructing the noisy ECG channel  is a  simplified version, but equally effective,  of the winning entry in that Challenge. We show that the procedure is robust against noise in the input signals and can include, as an input, the channel we want to denoise. 

This noise removal method is another example of the power of deep neural networks~\cite{hintonReducingdimension, hintonFastLearning, hintonRecognizeshapes, BengioLecun2007:scaling}, in this case, applied to ECG signals.

\section{Method}

\begin{figure*}[!ht]
  \centering
\includegraphics[height=8cm, width=13cm]{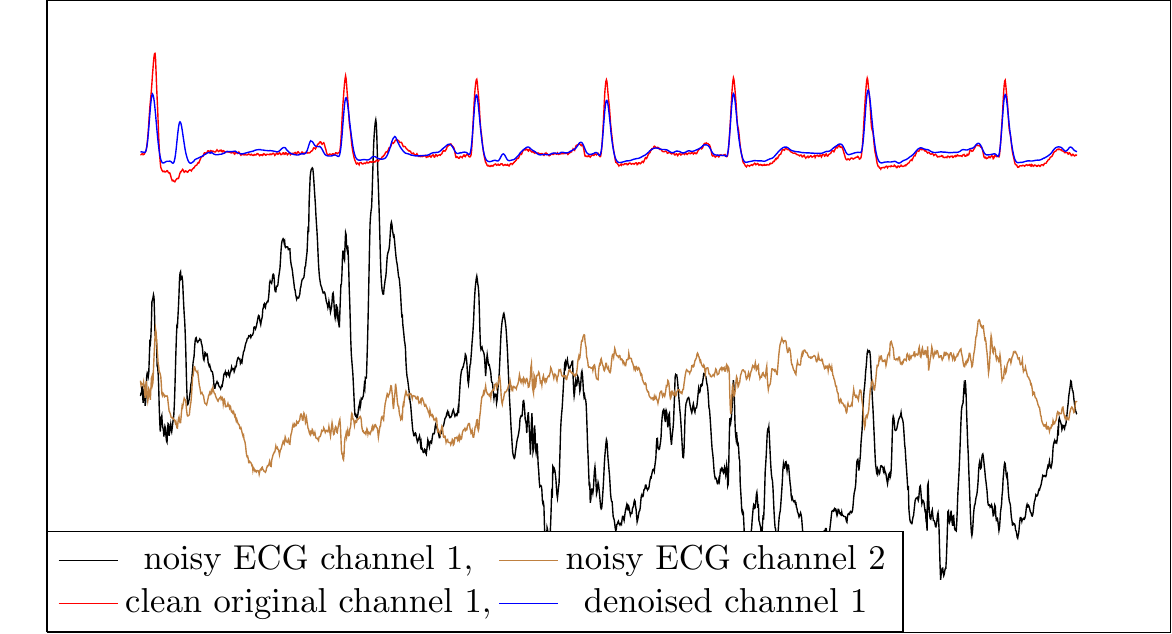}
\caption{Reconstructing the first channel from record 105 (MIT-BIH Arrhythmia Database), SNR=-6 db.  In the lower section, noisy ECG, in the
upper section, clean channel 1 and denoised channel 1.}
\label{fig:rec105}
\end{figure*}

 If  an ECG channel  we want to use for ECG analysis is, at some time segment,  contaminated with noise, we call it the {\bf target} channel in our denoising process. 

The method uses a feedforward neural network.
A prerequisite for applying it is the target channel to be free from noise for some minutes, in order to train the neural network. The other channels used by the procedure may have noise  in training time. If one channel has much more noise than others, it might be better not to use it for the reconstruction, even if it is the target channel.  

The neural network will receive the samples as input, from one  or more channels, corresponding to time segments with a fixed length. The output will be the samples from the target channel corresponding to the time segment used in the input. We used time segments with lengths of between one and three seconds, depending on  the channels we use for reconstruction: one second if we use the target channel and another channel, two seconds, if we do not use the target channel, and three seconds, if we only use the target channel.

To reconstruct one ECG channel we collected time segments, $T_k$ from the ECG each one starting 16 samples after the preceding segment. After obtaining the output of the neural network, corresponding to each of the $T_k$, the value of the reconstruction on sample $t_0$ will be the average value of the sample outputs corresponding to $t_0$, using all $T_k$ that contain $t_0$.

The proposed method could be  applied to a Holter record, reconstructing those time segments where an important level of noise  
is identified and using the remaining time of the Holter record for training.

\subsection{Neural network architecture and training}

We used a  neural network with three hidden layers. The number of  units on each of those layers was approximately 1000 in all experiments. 
To train the neural network, we constructed a sequence of time segments  each one starting five samples after the beginning of the previous one. There is no need to use fiducial points to create input data to the neural network.

 We applied Geoffrey Hinton's method~\cite{hintonroyalsocietyreview,hintonReducingdimension,hintonMultipleLayers} to learn the  neural network weights: following initialization using a stack of Restricted Boltzmann Machines, we applied backpropagation  algorithm to fine tune the weights.
For details on  the training procedure for Restricted Boltzmann Machines, we refer to Hinton~\cite{hintonrbmpracticalguide}.

As usual, when using feedforward neural networks, we normalized the input data, to accelerate the learning process. 
First we applied a moving average filter, with the window size equal to the sampling rate. Then we subtracted the result from the signal, thus reducing the baseline wander. In the output signal, instead of the moving average filter we applied a median filter: it is more effective in the removal of baseline wander. 
Finally, we scaled the output signal to  have unit variance and multiplied the input signals by the same scale factor.   

We implemented our method using  GNU Octave language and, to reduce training and reconstruction time, we ran most time consuming code on a Gra\-phics Processing Unit. Our code is available at the first author's web page.

\subsection{Evaluating the method}

\begin{figure*}[!ht]
  \centering
\includegraphics[height=8cm, width=13cm]{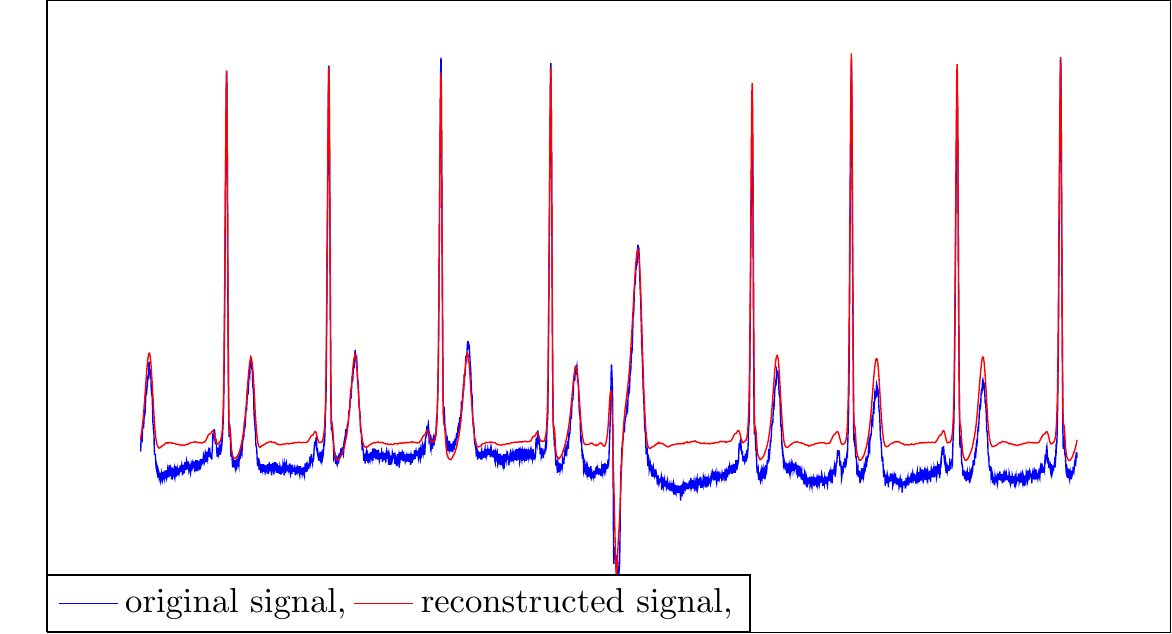}
\caption{Segment of first channel from record 202 (MIT-BIH Arrhythmia database): RMSE of recontructed signal is largely due to the noise in the original signal: baseline shift and high frequency noise.}
\label{rec202}
\end{figure*}

Evaluating ECG denoising methods is not an obvious task. A common way of doing it is to add add noise to an existing signal and measure the Root Mean Square Error (RMSE) of the denoised signal relative to the original signal. This approach has some disadvantages. Firstly, when using a large data base of ECGs, is difficult to avoid noise in the original signal, and we do not want to punish the denoising method for not reconstructing the noise in the original signal. Secondly,  RMSE does not always reflect the difficulties in analysing a noisy ECG. For instance, a constant baseline shift in the reconstructed signal is not very disturbing, but might correspond to a high RMSE value.

In this study we report RMSE in the reconstructed signal when we artificially add noise in the ECG, but we also evaluate our method using some publicly available programs that analyse the  ECG: we compare the results of applying these programs with and without denoising the corrupted ECG. Although those programs alredy  have a preprocessing stage to deal with noise, we show that,  in the presence of noise, our denoising method improves their results. 
 
\subsubsection{Programs used to test this method}

\begin{description}
\item[gqrs] is a recent QRS detector, not yet published: the author is George Moody. This program is open source and available with WFDB software,  from Physionet. There is an accompaining post-processor 'gqpost' intended to improve positive predictivity, at a cost of reduced sensitivity. We report the results of 'gqrs' with and without using gqpost. 'gqpost' uses a  configuration file 'gqrs.conf'; we kept the  default values of 'gqrs.conf'.  
 The results of this program 'gqrs' depend on the value of a threshold parameter; as we did not find a systematic way of determining, for each record, the best value for the threshold, we used the parameter's default value. For this reason, we do not  report the best possible results of this detector in the different records and therefore we should not use this study to compare the perfomance of the different qrs detectors. 

\item[E.P. limited ] is an open source program written by  Patrick S. Hamilton~\cite{Eplimitedqrs}. It performs QRS detection and classifies beats as 'normal' or 'ventricular ectopic' (VEB). 
\item[ecgpuwave] It is an open source QRS detector and waveform limit locator,   available as part of the PhysioToolkit~\cite{waveboudaries,evaluationwaveformlimits}. The authors are Pablo Laguna, Raimon Jané, Eudald Bogatell, and David Vigo Anglada. 
\end{description}

All the programs listed above act on a single ECG channel, we did not find publicly available methods using more than one channel.

\subsection{Statistics used to describe the results of QRS detectors and the beat classifier}
For QRS detectors we used the following statistics:
$$\mbox{Sensitivity}=\frac{TP}{TP+FN}\, ,\,\,\,\,\mbox{Positive Predictivity}=\frac{TP}{TP+FP}$$
$$\mbox{Error rate}=\frac{FP+FN}{TP+FN}\mbox{, \cite{Eplimitedqrs}}$$
 where $TP$ is the number of correctly detected beats, $FP$ is the number of false detections and $FP$ is the number of missed beats.
 For the beat classifier, we  use, as in~\cite{PatientAdaptable}, the Sensitivity, Positive Predictivity, 
$$\mbox{False Positive Rate}=\frac{FP}{TN+FP}$$
$$\mbox{and}\,\, \,\,\, \, \mbox{Classification Rate}=\frac{TN+TP}{TN+TP+FN+FP}$$
where $TP$, $TP$, $FP$ and $FN$ are defined as follows:
\begin{itemize}
\item $\mathbf{TP}$ is the number of beats correctly classified as VEB.
\item $\mathbf{TN}$ is the number of non VEBs correctly classified.
\item $\mathbf{FP}$  is the number of beats wrongly classified as VEB, excluding fusion and unclassifiable beats.
\item  $\mathbf{FN}$ is the number of true VEB not classified as such.
\end{itemize}
  
\subsection{Adding noise to an existing  ECG}

In most experiments,  to test the behavior of our denoising method, we start with a 'clean' ECG and add noise to it.
For this we use the program {\bf nst}, written by Geoge Moody~\cite{Moodynoisestress}.

 The standard definition of  signal to noise ratio (SNR), in decibels, is:
$$SNR=10\log_{10}\frac{S}{R}$$
where  $S$ and $R$ are the power of signal and noise.  We used a slightly different value for $S$ and $R$, following the method used by the program 'nst'.  Next we quote 'nst' man page~\cite{nstmanpage}:
\begin{quotation} \em
`` A  measurement  based  on  mean
       squared  amplitude,  for exam\-ple, will be proportional to the square of
       the heart rate.  Such a measurement  bears  little  relationship  to  a
       detector's  ability to locate QRS complexes, which is typically related
       to the size of the QRS complex.  A less  significant  problem  is  that
       unweighted  measurements  of noise power are likely to overestimate the
       importance of very low frequency noise, which is both common and  (usually)  not  troublesome  for  detectors.   In view of these issues, nst
       defines S as a function of the QRS amplitude, and N as a frequency-weighted noise power measurement. ``
\end{quotation}
More details on the way 'nst' computes SNR can be found on the man page of 'nst'.

\section{Experiments}

\subsection{MIT-BIH Arrhythmia Database}
\label{main_exper}

\begin{figure*}[!ht]
  \centering
\includegraphics[height=8cm, width=13cm]{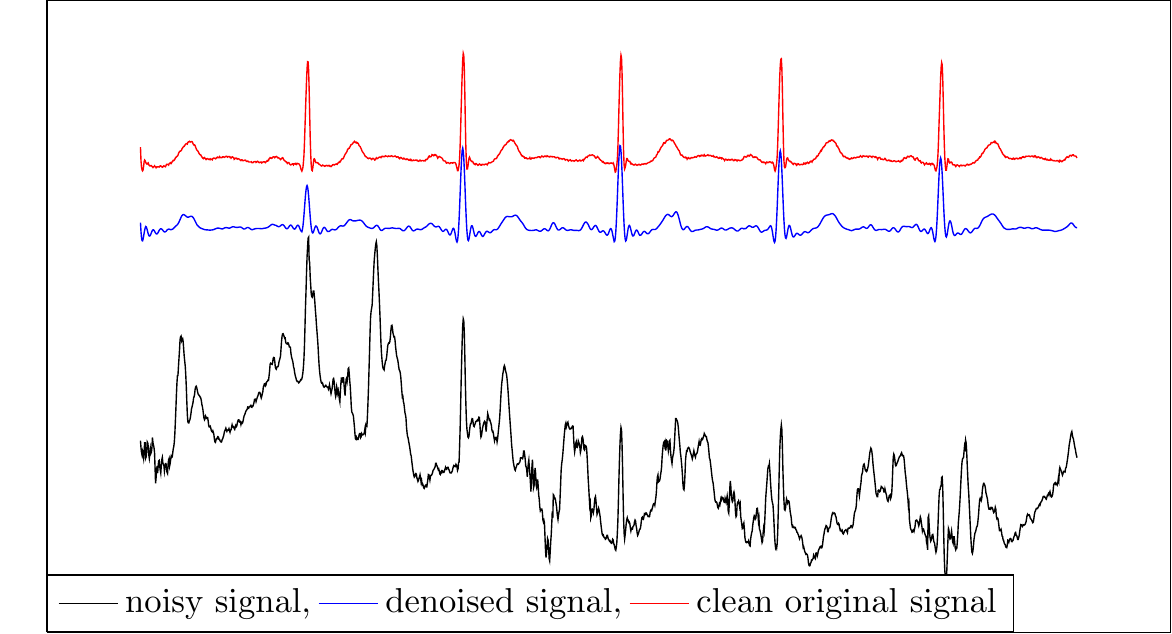}
\caption{Reconstructing channel 1 from record 103 using only the same channel, SNR=0 db.  At the bottom, noisy signal, in the middle denoised signal
 and at the top, clean signal.}
\label{fig:rec103}
\end{figure*}

We  added noise to both channels in all the 48 records from Physionet MIT-BIH Arrhythmia Database~\cite{mitbih-Arrhythmia, PhysioNet}, and applied our method to reconstruct the first channel. 
As it is well known~\cite{Moodynoisestress, similar}, from the three types of noise, baseline wander, muscle artifact and  electrode motion artifact, it is the last one that creates most difficulties to ECG analysis programs. We contaminated both channels of each record with electrode motion artifact noise, using the corresponding noise record from the MITBIH Noise Stress Test Database~\cite{Moodynoisestress}.

In all but one record, both channels were used as input to reconstruct the first channel.  In record 103, the noise in the second channel is already very high, therefore, we chose to use only the target channel in the reconstruction.

The clean record and corrupted noisy versions of the same record were used as input for training the neural network. We always used the clean target channel for the output.

The default behavior of the program 'nst' was followed to add noise to the records used in the tests: starting after the first five minutes from the beginning of each record, we added noise for two minutes, followed by two minutes without noise, repeating the process until the end of the record.

In order to train the neural network, we used all those segments of time where noise was not added in test time. In this way, the parts of  noise  used during training  and testing  do not overlap: we kept the neural network from learning the noise used in the test. The amount of noise, used for testing, corresponds to SNR values of  24 db, 18 db, 12 db, 6 db, 0 db and -6 db.

\begin{table*}[!ht]
\centering
 \caption{Reconstruction Error: for each level of noise, we report the value of RMSE(denoised signal)/RMSE(noisy signal) over the 48 records of MIT-BIH Arrhythmia Database}
\begin{tabular}[l]{c c c c c c c} \cline{1-7}
SNR&24db&18db&12db&6db&0db&-6db\\
    &1.014&0.507&0.257&0.133&0.071&0.041\\  \cline{1-7}
\end{tabular}
\label{RMSE}
\end{table*}

In table~\ref{RMSE} we report the fraction of RMSE, in the noisy signal, present in the reconstructed signal: RMSE(denoised signal)/RMSE(noisy signal).  As we can see in the table, there are no visible advantages, in terms of RMSE, in applying the denoising methods for low noise (SNR=24db), in fact, situations like the one in figure~\ref{rec202} introduce high values of RMSE because the method is not intended to learn to reproduce the noise of the original signal but just its main pattern. When the values of added noise increase, the errors in the reconstructed signal, due to noise in the original signal, lose their relative  importance: for higher values of noise we notice an important reduction in the value of RMSE in the reconstructed signal.As supplementary material to this article, we present the detailed results for each record. 

\begin{table*}[!ht]
\centering
 \caption{Results of 'gqrs' applied to the first channel, corrupted with noise and denoised using our method (MIT-BIH Arrhythmia, 48 records).}
\begin{tabular}[l]{c c c c c}
\multirow{2}{*}{SNR}& \multirow{2}{*}{channel 1}&\multirow{2}{*}{sensitivity}& positive& error\\ 
&  & &predictivity&rate \\
\cline{1-5}
\multicolumn{1}{c}{\multirow{2}{*}{24 db}}
 & noisy &  0.9973  & 0.9963& 0.0064\\

&\bf denoised&0.9963  &0.9992 & 0.0045\\ 

\cline{1-5}
\multicolumn{1}{c}{\multirow{2}{*}{18 db}}
 & noisy &  0.9973 &  0.9834 & 0.0195 \\

&\bf denoised& 0.9964  &  0.9991& 0.0045\\ 

\cline{1-5}
\multicolumn{1}{c}{\multirow{2}{*}{12 db}}
 & noisy &  0.9957  &  0.9054& 0.1084\\
&\bf denoised& 0.9959     &     0.9991& 0.0049\\ 

\cline{1-5}
\multicolumn{1}{c}{\multirow{2}{*}{6 db}}
 & noisy &  0.9881  &   0.7605& 0.3231\\

&\bf denoised& 0.9941 &  0.9989& 0.0071\\ 

\cline{1-5}

\multicolumn{1}{c}{\multirow{2}{*}{0 db}}
 & noisy & 0.9680     &     0.6471&   0.5599 \\ 
 &\bf denoised& 0.9826   & 0.9922&  0.0251 \\ 

\cline{1-5}

\multicolumn{1}{c}{\multirow{2}{*}{-6 db}}
 & noisy & 0.9523     &     0.5806&0.7357\\

 &\bf denoised&0.9470   &   0.9466 & 0.1064\\ 

\cline{1-5}

\end{tabular}
\label{tablegqrsmitbih}
\end{table*}

\begin{table*}[!ht]
\centering
 \caption{Results of 'gqrs' with post-processor 'gqpost' applied to the first channel, corrupted with noise and denoised using our method (MIT-BIH Arrhythmia, 48 records).}
\begin{tabular}[l]{c c c c c}
\multirow{2}{*}{SNR}& \multirow{2}{*}{channel 1} &\multirow{2}{*}{sensitivity}& positive&error\\ 
&  & &predictivity&rate \\
\cline{1-5}
\multicolumn{1}{c}{\multirow{2}{*}{24 db}}
 & noisy & 0.9970 & 0.9970 & 0.0060\\

&\bf denoised&  0.9961 & 0.9994& 0.0045\\ 

\cline{1-5}
\multicolumn{1}{c}{\multirow{2}{*}{18 db}}
 & noisy &   0.9969 & 0.9868 & 0.0165\\

&\bf denoised& 0.9960 & 0.9993 & 0.0047\\ 

\cline{1-5}
\multicolumn{1}{c}{\multirow{2}{*}{12 db}}
 & noisy & 0.9921 &  0.9318 & 0.0806\\

&\bf denoised& 0.9957 & 0.9992 & 0.0050\\ 

\cline{1-5}
\multicolumn{1}{c}{\multirow{2}{*}{6 db}}
 & noisy &  0.9659  & 0.8282 & 0.2345\\

&\bf denoised& 0.9942  & 0.9988& 0.0069\\ 

\cline{1-5}

\multicolumn{1}{c}{\multirow{2}{*}{0 db}}
 & noisy & 0.9121 & 0.7037& 0.4720\\ 
 &\bf denoised& 0.9838 & 0.9947&0.0215\\ 

\cline{1-5}

\multicolumn{1}{c}{\multirow{2}{*}{-6 db}}
 & noisy & 0.8767  & 0.6301 & 0.6380\\

 &\bf denoised&0.9493 & 0.9568 & 0.0935 \\ 

\cline{1-5}

\end{tabular}
\label{tablegqrsgqpmitbih}
\end{table*}

We applied the programs 'gqrs' and 'EPlimited' to the first channel, in noisy versions of each record and in the reconstructed signal, to verify whether, after applying our method, there were  significant improvements in the performance of those programs.  
The results are reported in tables~\ref{tablegqrsmitbih}, \ref{tablegqrsgqpmitbih}, \ref{tableEPLimitedmitbihqrs}, \ref{mitbihbeatclassification} . The  first column  indicates the SNR of the resulting ECG, the same value for both channels, after corrupting it with noise. The second column refers  to the  signal used  when applying the program to the first channel: {\bf denoised} means the reconstructed noisy first channel, using our method. The tables present the sensitivity, positive predictivity, number of detection errors and error rate, in the case of QRS detectors, and  VEB sensitivity, positive predictivity, false positive rate and classification rate, for the 'EP Limited' beats classification. We used the following programs to report the results:  'bxb', from WFDB software~\cite{standards}, in the case of 'gqrs', and 'bxbep', in the the case of 'EPLimited'.
 The numbers are relative to all the 48 records from the MIT-BIH  Arrhythmia Database, 91225 beats, from which 6101 are VEBs: we started the test after the first 5 minutes and stopped one second before the end. In the case of EP limited we had to start the test one second later because we could not configure 'bxbep'to behave  in another way.

\begin{table*}[!ht]
\centering
\caption{Results of EP Limited (QRS detection) applied to the first channel, corrupted with noise and denoised using our method(MIT-BIH Arrhythmia, 48 records). }
\begin{tabular}[l]{c c c c c} 
\multirow{2}{*}{SNR}& \multirow{2}{*}{channel 1}&\multirow{2}{*}{sensitivity}& positive&  error \\
 & & & predictivity&rate\\ \cline{1-5}

\multicolumn{1}{c}{\multirow{2}{*}{24 db}}
 & noisy & 0.9977  & 0.9981& 0.0042 \\ 
&\bf denoised&0.9961  &0.9996 & 0.0043\\ \cline{1-5}

\multicolumn{1}{c}{\multirow{2}{*}{18 db}}
 & noisy & 0.9977 & 0.9945& 0.0079 \\ 
&\bf denoised&0.9957 & 0.9995& 0.0048\\ \cline{1-5}

\multicolumn{1}{c}{\multirow{2}{*}{12 db}}
 & noisy & 0.9969 &0.9342& 0.0733 \\ 
&\bf denoised&0.9955 & 0.9995& 0.0050\\ \cline{1-5}

\multicolumn{1}{c}{\multirow{2}{*}{6 db}}
 & noisy& 0.9857 &0.7943& 0.2696 \\ 
&\bf denoised&0.9944 &0.9993&0.0063\\ \cline{1-5}

\multicolumn{1}{c}{\multirow{2}{*}{0 db}}
 & noisy &  0.9432& 0.7110& 0.4401\\ 
 &\bf denoised&0.9865 & 0.9951 &0.0184\\ \cline{1-5}

\multicolumn{1}{c}{\multirow{2}{*}{-6 db}}
 & noisy& 0.8557& 0.6568&0.5915 \\ 
 &\bf denoised&0.9531 &0.9493&0.0977\\ \cline{1-5}

\end{tabular}
\label{tableEPLimitedmitbihqrs}
\end{table*}

\begin{table*}[!ht]
\centering
\caption{Beat clasification as Normal or Ventricular Ectopic Beat (VEB): results of EP Limited applied to the first channel, corrupted with noise and denoised using our method(MIT-BIH Arrhythmia, 48 records). }
\begin{tabular}[l]{cccccc} 
\multirow{2}{*}{SNR}& \multirow{2}{*}{signal}& VEB & VEB positive& VEB false & classification \\
 & &sensitivity &  predictivity & positive rate& rate \\ \cline{1-6}
\multicolumn{1}{c}{\multirow{2}{*}{24 db}}
 & noisy&0.9147 & 0.9589 & 0.0028 & 0.9916\\
&denoised &0.9142 &  0.9815& 0.0013& 0.9930 \\ \cline{1-6}

\multicolumn{1}{c}{\multirow{2}{*}{18 db}}
 & noisy &  0.8873  & 0.9260 & 0.0051&0.9876\\
&denoised& 0.9089  &  0.9759&0.0016&0.9923\\ \cline{1-6}

\multicolumn{1}{c}{\multirow{2}{*}{12 db}}
 & noisy & 0.8190 &0.5935 & 0.0380& 0.9530\\
&denoised&0.9032 & 0.9785&0.0014&0.9921\\ \cline{1-6}

\multicolumn{1}{c}{\multirow{2}{*}{6 db}}
 & noisy & 0.6977&0.2369&0.1291&0.8615\\
&denoised&0.8901 &  0.9778&0.0015&0.9912\\ \cline{1-6}

\multicolumn{1}{c}{\multirow{2}{*}{0 db}}
 & noisy & 0.6083& 0.1308& 0.2160&0.7751\\
&denoised&0.8567 &  0.9513& 0.0032&0.9873\\ \cline{1-6}

\multicolumn{1}{c}{\multirow{2}{*}{-6 db}}
 & noisy & 0.5720& 0.0939& 0.2994&0.6940\\
&denoised&0.7663  &0.7689  &0.0166& 0.9688\\ \cline{1-6}
\end{tabular}
\label{mitbihbeatclassification}
\end{table*}

For QRS detectors, after applying our denoising procedure,  there is always an improvement in positive predictivity  in the tested programs, but, for high values of SNR there is a small reduction in the sensitivity: above 12 db for gqrs and above 6 db for EPLimited. Besides some ectopic beats  not being well reconstructed, the reduction in sensitivity is due to a smaller amplitude of the QRS complex in the reconstructed signal; this occurs in the first beat from  figure~\ref{fig:rec103}. We could improve the  sensitivity in the reconstructed signal, at the cost of a reduction in the positive predictivity, multiplying the reconstructed signal by a factor greater than 1.0, but we chose not to do it.

For beat classification there is always a clear improvement after using the proposed method. 
  
As supplementary material to this article, we present the detailed results for each program and record.

\subsection{Record mgh124}

The MGH/MF Waveform Database~\cite{mghdatabase, PhysioNet} is a collection of electronic recordings of hemodynamic and electrocardiographic waveforms. Typical records include three ECG leads. In the case of record mgh124, the first two ECG channels are sometimes strongly contaminated with noise, while the third ECG channel mantains a relatively good quality, therefore we have reliable QRS annotations. Using record mgh124, we tested our denoising method on a real ECG, without having to artificially add noise. In this case we reconstructed the second ECG channel, using  only that same channel as input:  we trained a neural network to produce a clean segment of the second channel given a corrupted version of the same segment. The clean parts of the  channel 2 were used to obtain training data for the neural network.

\begin{figure*}[!ht]
  \centering
\includegraphics[height=8cm, width=13cm]{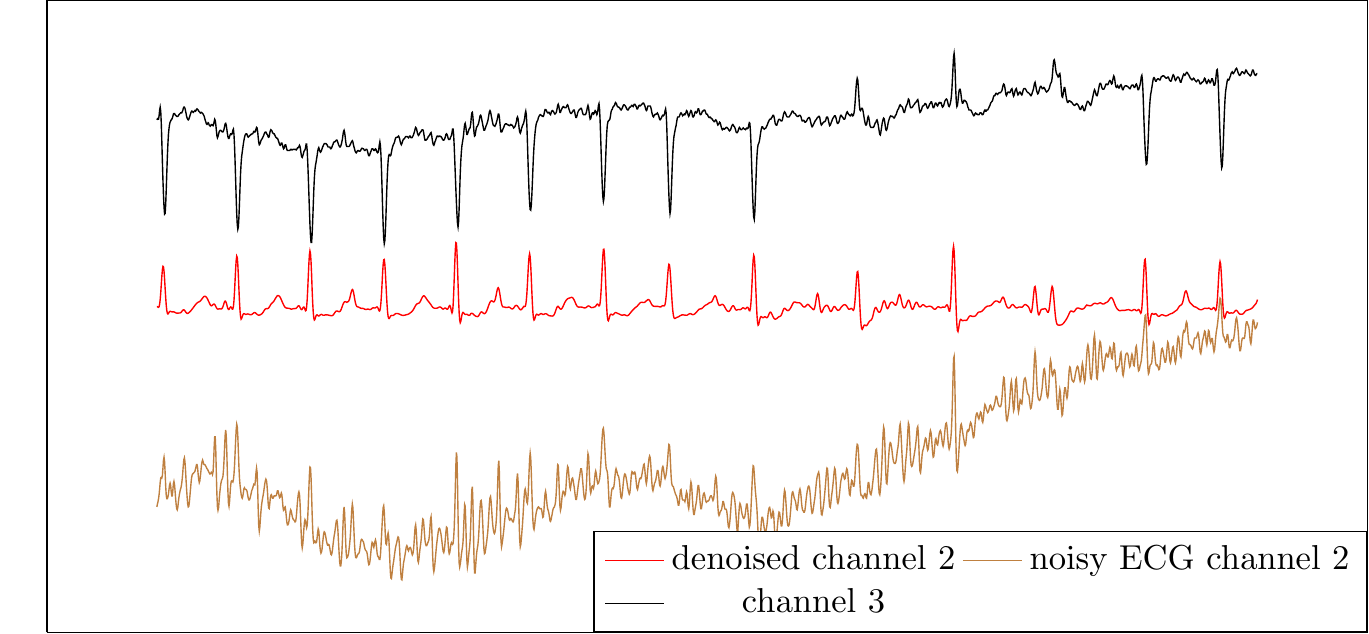}
\caption{Reconstruction of noisy channel 2 from record mgh124: channel 3 is shown as a reference, it is not used for denoising. We notice that although the first VEB is recognized as such, the next two are not. This ECG segment starts at sample 853318.}
\label{fig:recmgh124}
\end{figure*}

tables ~\ref{tablegqrsmgh124} and~\ref{tableEPLmgh124} show the results of 'gqrs' and 'EP Limited' on the original second ECG channel and on the reconstructed version. The total number of beats during testing time is 8573, from which 458 are classified as VEBs.

\begin{table*}[!ht]
\centering
 \caption{Results of 'gqrs' applied to second ECG channel, corrupted with noise and denoised using our method (record mgh124).}
\begin{tabular}[l]{ c c c c c}
\cline{1-5}
\multirow{2}{*}{lead 2}&  detection &error&\multirow{2}{*}{sensitivity}& positive\\ 
  &errors &rate& &predictivity \\
\cline{1-5}
 noisy &  1344  &  0.1568 & 85.06\% &  99.14\%\\
\bf denoised& 890&  0.1038& 0.8971 &0.9990\\ 
\cline{1-5}
\end{tabular}
\label{tablegqrsmgh124}
\end{table*}

\begin{table*}[!ht]
\centering
\caption{EP Limited: QRS detection and beat clasification as Normal or Ventricular Ectopic Beat (VEB), results on lead 2, corrupted with noise and denoised using our method(record mgh124). }
\begin{tabular}[l]{ccccc} 
\multirow{2}{*}{signal}&   QRS detection & QRS error& \multirow{2}{*}{ QRS sensitivity}& QRS positive \\
  &errors &rate&  &predictivity\\ \cline{1-5}
noisy & 553&0.0645  &0.9472& 0.9878\\ 

 \bf denoised& 463 &0.0540& 0.9470 &0.9989\\ \cline{1-5}
& & &\\ \cline{1-4}
\multirow{2}{*}{signal}&  \multirow{2}{*}{VEB sensitivity}& VEB positive& VEB false\\
& & predictivity & positive rate \\\cline{1-4}
 noisy & 23.14\% & 17.82\%&6.305\%\\
recsignal&36.68\%&87.50\%&0.311\%\\ \cline{1-4}
\end{tabular}
\label{tableEPLmgh124}
\end{table*}

\subsection{record sele0106 from QT database}

Determination of peaks and limits of ECG waves is very important for ECG analysis: they are necessary for ECG measurements that are clinically relevant, namely, PQ interval, QRS duration, ST segment and QT interval.

Physionet QT data base was created to test QT detection algorithms~\cite{QTdatabase}. Each record contains at least 30 beats with manual annotations identifying the beginning, peak and end of the P-wave, the beginning and end of the QRS-complex and the peak and end of the T-wave.

 We used the program 'ecgpuwave' to show that, in some situations, we can improve automatic ECG delineation by using a clean channel to reconstruct a very  noisy one.

Typically, the accuracy of  ecgpuwave when detecting the limits or peak of some ECG characteristic wave is better in one of the channels. The best channel to locate one of those  reference points changes with the different characteristic points and also from record to record. 

Table~\ref{sele0106:clean leads} shows the results of
ecgpuwave, on the two channels, when it locates P wave peak, P ending, QRS beginning and QRS ending. We are using the first annotator as reference. We can conclude that the error is smaller when ecgpuwave is applied to the second channel.

\begin{table*}[!ht]
\centering
\caption{Accuracy of ecgpuwave, in ms, locating some characteristic points on the record sele0106(QT database): comparing results for the two leads.}
\begin{tabular}[l]{cccc}
\cline{1-4}
reference point& channel& mean error & std error\\ \cline{1-4}
\multirow{2}{*}{P peak} & 1& 9.33 & 4.54\\ 
 & 2 & 2.93 & 2.72\\ \cline{1-4}
\multirow{2}{*}{P off} & 1 & 15.73 & 10.06 \\
 & 2 & 7.47 & 9.28\\ \cline{1-4}
\multirow{2}{*}{QRS on} & 1 &  18.80 & 6.62\\
 & 2&  7.87  &  4.32  \\ \cline{1-4}
\multirow{2}{*}{QRS off} & 1& 13.33 & 4.17\\
 & 2 & 4.67 &  4.27 \\ \cline{1-4}
\cline{1-4}
\end{tabular}
\label{sele0106:clean leads}
\end{table*}

At this point we consider an easily imaginable situation, where the second channel is highly corrupted with noise, in such a way that it is better to use only the first channel for the reconstruction of the second channel. In this case we trained a neural network to produce a segment of channel 2 when it gets  the corresponding segment of channel 1 as input. 

\begin{figure*}[!ht]
  \centering
\includegraphics[height=6cm, width=12cm]{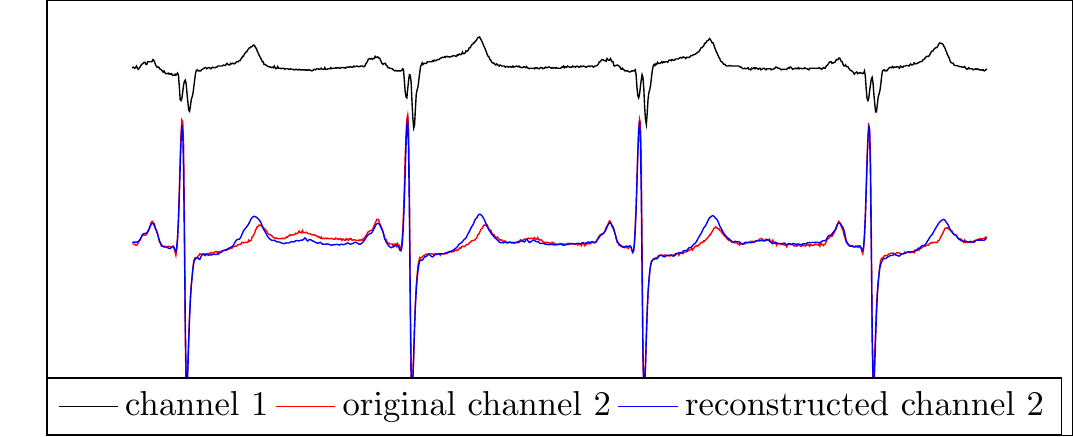}
\caption{reconstructing channel 2 from record sele0106(QT database) using only channel 1.}
\label{fig:recsele0106}
\end{figure*}

 We followed this procedure and applied ecpuwave to the reconstructed channel~2. The results are in table~\ref{sele0106:reclead2}.
One can see that we still get better results using reconstructed channel~2, from channel 1, than when applying ecgpuwave to clean channel~1.

\begin{table*}[!ht]
\centering
\caption{comparing results, in ms, of ecgpuwave for the channel 1 and reconstructed channel 2}
\begin{tabular}[l]{cccc}
\cline{1-4}
reference point& channel& mean error & std error\\ \cline{1-4}
\multirow{2}{*}{P peak} & 1& 9.33 & 4.54\\ 
 & reconstructed channel 2 & 2.80 & 2.76\\ \cline{1-4}
\multirow{2}{*}{P off} & 1 & 15.73 & 10.06 \\
 & reconstructed channel 2 & 7.20 & 6.54\\ \cline{1-4}
\multirow{2}{*}{QRS on} & 1 &  18.80 & 6.62\\
 & reconstructed channel 2&  7.33  &  4.63  \\ \cline{1-4}
\multirow{2}{*}{QRS off} & 1& 13.33 & 4.17\\
 & reconstructed channel 2 & 6.40 &  4.21 \\ \cline{1-4}
\cline{1-4}
\end{tabular}
\label{sele0106:reclead2}
\end{table*}

\section{Discussion of results and conclusions}

Adding  noise to existing records, we  carried out extensive expe\-ri\-ments on all the records from the MIT-BIH Arrhythmia Database. In the presence of high noise, SNR equal to 12db and lower, the programs we tested showed much better perfomance when we  applied our denoising method to the ECGs. For low  noise, SNR above 12db, after applying our method, QRS detectors show a slight reduction in sensitivity  although there is an improvement in the positive predictivity. 
 The experiments with records mgh124 and sele0106, without artificially adding noise in the test, confirm the advantages of using our method on a real ECG,  a Holter record, for example.  The experiment with record sele0106 also shows that the result of reconstructing a noisy channel can be exceptionally good when clean channels are available.

\bibliographystyle{plain} 

\end{document}